\documentclass[twocolumn,showpacs,prb,amsmath,amssymb,superscriptaddress]
{revtex4}
\usepackage{graphicx}
\usepackage{bm}
\usepackage{color}
\DeclareMathOperator{\Tr}{Tr} 

\DeclareMathOperator{\Max}{Max}

\DeclareMathOperator{\Sgn}{sgn}

\begin{document}
\title{Microscopic model of critical current noise in
  Josephson-junction qubits: Subgap resonances and Andreev bound states}
\author{Rog\'{e}rio de Sousa} \affiliation{Department of Physics and
  Astronomy, University of Victoria, Victoria, BC V8V 4H3, Canada}
\author{K. Birgitta Whaley} \affiliation{Department of Chemistry and
  Pitzer Center for Theoretical Chemistry, University of California,
  Berkeley, California 94720-1460, USA} 
\author{Theresa Hecht}
\affiliation{Physik Department, CeNS, and ASC,
  Ludwig-Maximilians-Universit\"at, Theresienstr. 37, D-80333
  M\"unchen, Germany} 
\author{Jan von Delft} 
\affiliation{Physik
  Department, CeNS, and ASC, Ludwig-Maximilians-Universit\"at,
  Theresienstr. 37, D-80333 M\"unchen, Germany} 
\author{Frank K. Wilhelm}
\affiliation{Department of Physics and Astronomy and Institute for
  Quantum Computing, University of Waterloo, 200 University Avenue W,
  Waterloo, ON, Canada, N2L 3G1} 
\date{\today}

\begin{abstract}
  We propose a microscopic model of critical current noise in
  Josephson-junctions based on individual trapping-centers in the
  tunnel barrier hybridized with electrons in the superconducting
  leads. We calculate the noise exactly in the limit of no on-site
  Coulomb repulsion. Our result reveals a noise spectrum that is
  dramatically different from the usual Lorentzian assumed in simple
  models. We show that the noise is dominated by sharp subgap
  resonances associated to the formation of pairs of Andreev bound
  states, thus providing a possible explanation for the spurious
  two-level systems (microresonators) observed in Josephson junction
  qubits [R.W.  Simmonds {\it et al.}, \prl {\bf 93}, 077003 (2004)].
  Another implication of our model is that each trapping-center will
  contribute a sharp dielectric resonance only in the superconducting
  phase, providing an effective way to validate our results
  experimentally.  We derive an effective Hamiltonian for a qubit
  interacting with Andreev bound states, establishing a direct
  connection between phenomenological models and the microscopic
  parameters of a Fermionic bath.
\end{abstract}
\pacs{
74.50.+r 
74.40.+k 
%
}
\maketitle

\section{Introduction}

\begin{figure}
\includegraphics[width=3in]{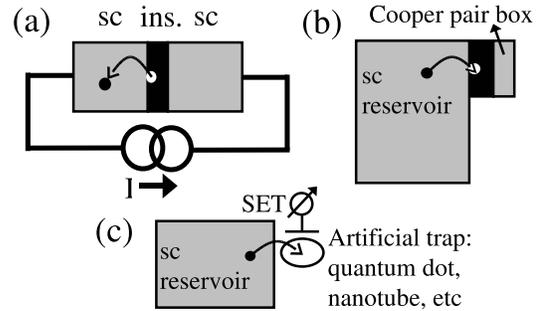}
\caption{(a) A current biased Josephson junction adversely 
  affected by the tunneling of electrons between one of the
  superconducting leads (sc) and a trapping-center defect in the insulating
  barrier (ins.).  Josephson junction critical current noise is directly
  related to fluctuations in trapping-center occupation due to
  modulation of the tunneling rate between superconducting leads.  (b)
  A Cooper-pair box quantum bit affected by charge noise produced
  by a single trapping-center in the barrier. 
  (c) Proposed set-up for measuring trap
  noise close to the superconducting transition. A single electron
  transistor (SET) is weakly coupled to an artificial trap, e.g. a
  normal state quantum dot or a nanotube.\label{noise_setup}}
\end{figure}

The performance of Josephson-junction devices functioning as units of
quantum memory or as qubits depends to a large extent on the amount of
charge and critical current noise affecting each Josephson-junction.
\cite{vanharlingen04,wellstood04,ithier05} One mechanism for critical
current noise is to assume that trapping-centers (TCs) located in the
tunnel barrier will partially block conduction whenever they capture
electrons from one of the superconducting electrodes
[Fig.~\ref{noise_setup}(a)].\cite{wakai87} The noise resulting from
each TC is traditionally modeled as two-level telegraph noise, with a 
Lorentzian noise spectrum, and a combination of several TCs
leads to $1/f$ noise.\cite{kogan96}

Nevertheless, sensitive spectroscopy experiments on current-biased
Josephson-junctions (phase qubits) revealed the presence of a few
microwave resonators on top of the expected $1/f$
noise.\cite{simmonds04} These microresonators behave as spurious
two-level systems buried within the tunnel barrier, whose coupling to
the qubit produces reduced measurement fidelity and
decoherence.\cite{cooper04} Similar microresonators were observed in
flux qubits.\cite{plourde05} The microscopic origin of the
microresonator remains unknown. However, there is strong evidence that
improving the junction oxide quality reduces their
concentration.\cite{oh06} Recently, phenomenological models based on
resonant coupling with the Josephson energy,\cite{simmonds04} and
dielectric resonance\cite{martinis05} were proposed.  Two measurement
schemes to distinguish these different models were
suggested.\cite{martin05,tian07} An interesting connection between the
low and high frequency scales of the noise spectrum due to a large
number of microresonators was demonstrated.\cite{shnirman05} To our
knowledge, there are two proposals in the literature for the
microscopic origin of these microresonators.  The first is based on
macroscopic resonant tunneling in large
Josephson-junctions.\cite{johnson05} This model explains the splitting
of the Josephson energy but predicts no dielectric resonance for the
microresonator.  The second microscopic model is based on the
structural two-level system in glasses.\cite{martinis05,constantin07}
This gives rise to the \emph{same} dielectric resonance above and
below the superconducting critical temperature.  Recently, a quantum
computer architecture using microresonators as qubits was
proposed.\cite{zagoskin06} Therefore, understanding the microscopic
origin of the microresonator is of central importance for improving superconducting qubits.

In a previous letter, we studied the charge noise spectrum due to a
single TC hybridized with a non-superconducting Fermi
sea.\cite{desousa05} At high temperatures we showed that the presence
of a single TC with energy level close to the Fermi level leads to the
expected Lorentzian spectrum characteristic of semiclassical random
telegraph noise.  At lower temperatures and frequencies below the TC
linewidth, the noise has a quantum Johnson-Nyquist form reflecting the
electron-hole excitations in the gate electrode Fermi
sea.\cite{desousa05}

Here we consider the noise spectrum due to a single TC hybridized with
electrons in a superconductor.  We show that the noise spectrum of
each TC is characterized by a sharp resonance associated to the
Andreev bound states formed from the TC hybridization with the
superconductor.  We further demonstrate that our theory describes a
direct connection between this TC physics and the spurious
microresonators observed at subgap frequencies in Josephson-junction
devices.\cite{simmonds04,cooper04,plourde05} We derive an effective
Hamiltonian connecting the discrete levels to the microscopic
parameters of the TC plus superconductor bath. Since the TC has an
electric dipole moment, our model predicts that a sharp dielectric
resonance will appear when the sample becomes superconducting.

The model proposed here is based on tunneling events between
\emph{individual} TCs and the superconductor. This is different
from other models\cite{faoro05,faoro06,lutchyn08} that considered
charge tunneling between two TCs mediated by Andreev states, resulting
in a smooth noise spectrum that does not give rise to microresonators.

Trapping-center fluctuation nearby to single electron tunneling
devices\cite{zorin96} is also an important source of charge noise and
decoherence of charge qubits such as the Cooper-pair
box\cite{nakamura02,astafiev04} [Fig.~\ref{noise_setup}(b)] and double
quantum dot.\cite{hayashi03}

Fig.~\ref{noise_setup}(c) suggests a test device to probe TC noise
around the superconducting transition temperature $T_c$ that 
allows verification of our predictions in a controlled manner.  A tunable
artificial trap, which can be realized by a quantum dot or a nanotube
in the normal state,\cite{buitelaar02} is coupled to a large metallic
reservoir at temperatures close to the superconducting transition. A
single electron transistor (SET) is proposed to measure TC charge
occupation in real time.\cite{lu03} This will  map the emergence of
the subgap resonance as $T$ is lowered below the superconducting
transition temperature $T_c$.

\section{Qubit decoherence and quantum noise\label{section_qubit}}

The behavior of superconducting circuits containing
Josephson-junctions is markedly quantum-mechanical.  Hence one can
design circuits that behave as artificial two-level systems, realizing
promising qubits for scalable quantum computer
architectures.\cite{clarke08}.

Consider a model for an artificial two-level system,
\begin{equation}
{\cal H}_{\rm{Qubit}}=\frac{1}{2}\hbar\bm{\Omega}(\hat{I}_c)\cdot \hat{\bm{\sigma}},
\label{qubitH}
\end{equation}
where
$\hat{\bm{\sigma}}=(\hat{\sigma}_x,\hat{\sigma}_y,\hat{\sigma}_z)$ is
the vector of Pauli matrices denoting the qubit, and $\bm{\Omega}$ is
a vector with dimensions of frequency.  The latter is a function of
$\hat{I}_c$, the critical current of one of the Josephson-junctions in
the circuit.  We assume the critical current depends on the quantum
state of TCs in the barrier, hence we write it as an operator
(notation $\hat{I}_c$ to distinguish quantum operators from c-numbers
such as $I_c$).  For small fluctuations we may write
$\bm{\Omega}(\hat{I}_c)\approx
\bm{\Omega}_0+\bm{\Omega}_{0}^{'}(\delta \hat{I}_c)+{\cal O}(\delta
\hat{I}_c)^2$, where $\bm{\Omega}_0=\bm{\Omega}(\langle
\hat{I}_c\rangle)$ and $(\delta\hat{I}_c)=\hat{I}_c-\langle
\hat{I}_c\rangle$. Choosing a coordinate system with $z$-axis along
$\bm{\Omega}_0$, and $x$-axis along
$\bm{\Omega}_1=\bm{\Omega}_{0}^{'}-(\bm{\Omega}_{0}^{'}\cdot
\hat{z})\hat{z}$ we get
\begin{equation}
{\cal H}_{\rm{Qubit}}=\frac{1}{2}\hbar\Omega_0 \hat{\sigma}_z + \hbar\eta_{z}(\delta\hat{I}_c) \hat{\sigma}_z + 
\hbar\eta_{x}(\delta\hat{I}_c) \hat{\sigma}_x.
\label{qubitHtrans}
\end{equation}
Fluctuations in $\delta \hat{I}_c$ affects the
qubit through the parameters
$\eta_z=\frac{1}{2}\bm{\Omega}_{0}^{'}\cdot \hat{z}$ and $\eta_x =
\frac{1}{2}\Omega_1$. The former leads to phase relaxation or 
decoherence, while the latter causes energy relaxation. 

In the weak coupling regime, all relaxation effects are fully characterized by the critical current noise spectrum,
\begin{equation}
\tilde{S}_{I}(\omega)=\int_{-\infty}^{\infty}\frac{dt}{2\pi}
\textrm{e}^{i\omega t}
\left\langle \left[\hat{I}_{c}(t)-\langle \hat{I}_c\rangle\right]
\left[\hat{I}_{c}(0)-\langle \hat{I}_c\rangle\right]
\right\rangle.
\label{defNoise}
\end{equation}
For example, if the qubit is prepared in the excited state
($\mid\uparrow\rangle$), its rate of approach towards thermal
equilibrium will be given by
\begin{equation}
\frac{1}{T_1}=\frac{\pi}{2}\eta_{x}^{2}\left[\tilde{S}(\Omega_0)+\tilde{S}(-\Omega_0)\right].
\end{equation}
Similarly, if the qubit is prepared in a superposition state
$(\mid\uparrow\rangle+\mid\downarrow\rangle)/\sqrt{2}$, its coherence
envelope $\langle\hat{\sigma}_+\rangle=\langle \hat{\sigma}_x
+i\sigma_y\rangle$ will be affected by low frequency noise according
to\cite{martinis03,desousa06}
\begin{equation}
\left|\langle \sigma_{+}(t)\rangle\right|
=\exp{\left[-\eta_{z}^{2}\int_{-\infty}^{\infty}d\omega\;\tilde{S}_{I}(\omega)
{\cal F}(t,\omega)\right]}.
\label{splus}
\end{equation}
Here the filter function ${\cal F}(t,\omega)$ depends on the
particular method chosen for probing qubit
coherence. For free induction decay we have
\begin{equation}
{\cal F}_{\rm{FID}}(t,\omega)=
\frac{1}{2}\frac{\sin^2{\left(\omega t/2\right)}}{\left(\omega/2\right)^2}, 
\label{ffid}
\end{equation}
while for the Hahn echo
\begin{equation}
{\cal F}_{\rm{Hahn}}(2t_{e},\omega)=
\frac{1}{2}\frac{\sin^4{\left(\omega t_{e}/2\right)}}{\left(\omega/4\right)^2},
\label{fecho}
\end{equation}
with qubit coherence probed at $t=2t_{e}$ after the application of a
$\pi$-pulse at time $t_{e}$.  Note that ${\cal
  F}_{\rm{Hahn}}(2t_{e},0)=0$. The Hahn echo filters out terms
proportional to $\tilde{S}_{I}(0)$ hence leading to
much longer coherence times for qubits subject to low frequency
noise (See Ref.~[\onlinecite{desousa06}] for
further discussion and derivations). 

For the purposes of this work, it is instructive to use
Eq.~(\ref{splus}) to study the effect of a sharp frequency 
peak (a resonance) in the noise spectrum. Assume $\tilde{S}_{I}(\omega)$ 
has a sharp peak centered at $\Omega_{\rm{Res}}$ with linewidth $1/\tau_{d}$,
\begin{equation}
\tilde{S}_{I}(\omega)=\frac{\tau_d}{\pi}
\frac{1}{
\left(\omega-\Omega_{\rm{Res}}\right)^2
\tau_{d}^{2}+1}.
\end{equation}
Using Eqs.~(\ref{splus}) and (\ref{ffid}) 
and assuming $\Omega_{\rm{Res}}\gg 1/\tau_d$ we get
\begin{equation}
\left|\langle \sigma_+(t)\rangle\right| \approx 
\exp{\left[-2\left(\frac{\eta_z}{\Omega_{\rm{Res}}}\right)^2
  \left(1-\textrm{e}^{-t/\tau_d}\cos{\Omega_{\rm{Res}} t}\right)\right]}.
\label{visib}
\end{equation}
Therefore a resonance in the noise spectrum 
leads to loss of visibility of coherence oscillations. The loss of visibility is initially
oscillatory, but decays exponentially to a fixed contrast for $t\gg
\tau_d$, similar to [\onlinecite{wilhelm08}].  Although Eq.~(\ref{visib}) was calculated
for free induction decay, it is also a good approximation for Hahn
echoes in the limit $\Omega_{\rm{Res}}\gg 1/t_{e}$.

The above discussion makes clear the fact that the key quantity to be
studied in the context of qubit relaxation and decoherence is the time
ordered noise spectrum defined by Eq.~(\ref{defNoise}). If noise is
the object of interest, the qubit acts as a spectrometer for quantum
noise.\cite{schoelkopf02} Later, in section~\ref{sectioneffh} we are
going to show that the same basic Hamiltonian also leads to the
formation of avoided crossing with Andreev levels acting as junction
resonators.

\section{Microscopic model for critical current noise}

\subsection{Trapping-center model Hamiltonian}

The Hamiltonian for a trapping-center coupled to a lead with 
Bardeen-Cooper-Schrieffer (BCS) 
interactions is given by\cite{anderson61,tinkham04}
\begin{equation}
{\cal H}={\cal H}_{0}+ {\cal H}_{\rm{BCS}}+{\cal V}.
\label{htot}
\end{equation}
The unperturbed trap Hamiltonian reads
\begin{equation}
{\cal H}_{0}=\sum_{\sigma}\epsilon_{d} n_{\sigma},
\label{h00}
\end{equation}
where $n_{\sigma}=d^{\dag}_{\sigma}d_{\sigma}$ is the 
electron number operator for
a TC with spin $\sigma=\uparrow,\downarrow$, and $d^{\dag}_{\sigma}$
is a Fermion creation operator.  The trap energy level $\epsilon_{d}$
is measured with respect to the Fermi level (we assume
$\epsilon_F=0$). The unperturbed mean-field Hamiltonian for a
superconducting Fermi lead is given by
\begin{eqnarray}
{\cal H}_{\rm{BCS}}=\sum_{k,\sigma}\epsilon_k n_{k\sigma}
-\sum_{k}\Delta c^{\dag}_{k\uparrow}
c^{\dag}_{-k\downarrow}+ \rm{H.c.},
\label{hbcs}
\end{eqnarray}
where $c^{\dag}_{k\sigma}$ creates a conduction electron in the gate
electrode with energy $\epsilon_k$, and
$n_{k\sigma}=c^{\dag}_{k\sigma}c_{k\sigma}$. $\Delta$ is the
superconducting order parameter. 
The conduction electrons
are hybridized with the TC via the hopping Hamiltonian
\begin{equation}
{\cal V}=\sum_{k,\sigma}V_{k}
d^{\dag}_{\sigma}c_{k\sigma}+\rm{H.c.},
\label{vdk}
\end{equation}
where $V_{k}$ is the tunneling matrix element for 
the electron between the TC and the superconducting lead.

Here we assume TCs for which the on-site Coulomb repulsion, of the
form $U n_{\uparrow}n_{\downarrow}$, can be neglected.  We remark that
the chemical structure of TCs in the Josephson barrier is not known.
There are many possible kinds of TCs associated with the amorphous oxide
in a typical Josephson junction: O-H complexes, various kinds of
vacancies, dangling bonds, etc.  Our model will be applicable to TCs
with $U\ll \Delta$.  The $U=0$ idealization is an important starting
point, \emph{particularly because it allows an exact solution of the
  noise problem}.  As we show below, our model seems to explain some
of the important features observed in spectroscopy of Josephson
qubits. In section \ref{discussionandconclusions} we discuss the
expected modifications when $U>0$.

\subsection{Trapping-center fluctuation as a mechanism for
  critical current noise \label{cricurrsection}}

We now describe a model for the effect of TC fluctuation on the
critical current of a Josephson-junction [Fig.~\ref{noise_setup}(a)].
Our aim is to establish a direct relationship between critical current
noise and the TC noise spectrum:
\begin{equation}
\tilde{S}_{n}(\omega)=\int_{-\infty}^{\infty}\frac{dt}{2\pi}\textrm{e}^{i\omega
t} \left\langle \left[\hat{n}(t)-2\bar{n}
\right]\left[\hat{n}(0)-2\bar{n}\right]\right\rangle.
\label{tcnoise}
\end{equation}
Here $\hat{n}=\sum_{\sigma} d^{\dag}_{\sigma}d_{\sigma}$ is the total
number operator for electrons occupying the TC level.  The notation
$\langle\hat{A}\rangle=\Tr{\{\hat{\rho}_G\hat{A}\}}$ denotes
grand-canonical averages using the density operator
$\hat{\rho}_G=\textrm{e}^{-\beta({\cal H}-\mu N)}/Z_G$, with
$\beta=1/k_B T$ and $Z_G$ the grand canonical partition function.  In
the absence of a magnetic field,
$\langle\hat{n}_{\uparrow}\rangle=\langle\hat{n}_{\downarrow}\rangle\equiv\bar{n}$,
therefore we write $\langle\hat{n}\rangle= 2\bar{n}$ to simplify the
notation.

The presence of a TC will produce weak modulations on the junction
potential barrier.\cite{wakai87,vanharlingen04,constantin07} Our model
is to assume that the channel average matrix element for electrons
tunneling from one lead to the other depends on $\hat{n}$ according to
\begin{equation}
\hat{T}_{LR}\approx T^{(0)}_{LR} + T^{(1)}_{LR} \hat{n}.
\label{tlr}
\end{equation}
The critical current $\hat{I}_c$
[or equivalently, the Josephson energy $\hat{E}_J =(\hbar/2e)
\hat{I}_c$] is proportional to the modulus squared of
Eq.~(\ref{tlr}),\cite{josephson69} so that in the adiabatic limit, for
frequencies smaller than the inverse tunneling time,
\cite{ambegaokar82} this directly translates into a fluctuation of
the critical current
\begin{equation}
\hat{I}_c \approx I_{0c} \left(
1+ \frac{|T^{(1)}_{LR}|}{|T^{(0)}_{LR}|} \hat{n} \right).
\label{ic}
\end{equation}
The critical current noise is therefore given by 
\begin{eqnarray}
\tilde{S}_{I}(\omega)&=&
\frac{I^{2}_{c0}|T^{(1)}_{LR}|^{2}}{|T^{(0)}_{LR}|^2}
\int_{-\infty}^{\infty}\frac{dt}{2\pi}\textrm{e}^{i\omega
t} \left\langle \left[\hat{n}(t)-2\bar{n}
\right]\left[\hat{n}(0)-2\bar{n}\right]\right\rangle\nonumber\\
&=&(\delta I_c)^2 \tilde{S}_{n}(\omega).
\label{relIcIn}
\end{eqnarray}

Hence within the linear approximation [Eq.~(\ref{tlr})] the resulting
critical current noise is directly proportional to the TC charge
noise, $\tilde{S}_n(\omega)$. The proportionality constant can be
extracted directly from experiments probing critical current
noise.\cite{wellstood04,eroms06,wakai87} Below we focus on theoretical
calculations of the TC noise spectrum $\tilde{S}_n(\omega)$ under
different parameter regimes.

Our model assumes the TC is coupled to only one of the superconducting
leads. Within the one-lead approximation critical current modulations
are assumed to occur only through variations of inter-lead tunneling
due to population/depopulation of the trap [Eq.~(\ref{tlr})]. We
therefore neglect the possibility for the trap electron to enter
through one lead and exit through the other.  These processes will
lead to interesting phase dependent effects in the Josephson
current.\cite{glazman89,choi04} We are not aware of studies of
critical current noise in this regime. Nevertheless, for zero phase
difference between the leads, we may map the problem into a TC coupled
to a single lead.\cite{choi04} Therefore our results should remain
valid in this case provided the phase is set to zero. In an
experimental sample containing a few TCs we should expect that some of
these are coupled to a single lead, others are coupled to both leads.
The former case will lead to phase-independent noise, while the latter
is expected to generate a phase dependent noise spectrum. In this
context the theory developed here should be compared to measurements
of the \emph{phase independent contributions to critical current
  noise}.\cite{wellstood04} Note that the TC only couples to both
leads if it is in the middle of the junction with a difference in
separations to either lead being smaller than a tunnel length.  Given
that junctions are typically much thicker than a tunneling length to
the extent that the latter are known,\cite{heinrich08} the present
theory covers most of the possible TC locations.

In this work we calculate the noise spectrum
under the assumption that the TC remains in thermal equilibrium
with the superconducting reservoir. Therefore our results are valid at
the regime where non-equilibrium effects are weak or can be neglected.
This is the case for a current-biased Josephson-junction in the zero
voltage state, or whenever the voltage is low enough so that the
electrons in the lead may still be characterized by a Fermi
distribution.  The thermal equilibrium assumption implies that the
noise spectrum satisfies the detailed balance condition,
$\tilde{S}(-\omega)
=\textrm{e}^{-\frac{\hbar\omega}{k_{B} T}}\tilde{S}(\omega)$.
The finite frequency noise spectrum measured by a particular
detector depends on details such as the detector temperature $T_{D}$
(not necessarily equal to the TC plus Fermi sea temperature $T$). For
example, current noise measured by an LC circuit relates to our
calculated time ordered noise [Eq.~(\ref{relIcIn})] in the following
way\cite{lesovik97}
\begin{equation}
\tilde{S}^{\rm{(LC)}}_{I}(\omega)=K\left\{
\tilde{S}_{I}(-\omega)+\frac{1}{\textrm{e}^{\frac{\hbar\omega}
{k_B T_D}}-1}\left[\tilde{S}_{I}(-\omega)-\tilde{S}_{I}(\omega)
\right]\right\},
\label{lesovikexp}
\end{equation}
where $K$ denotes the effective coupling constant between the current
carrying wire and the LC circuit. The experiment proposed in
Fig.~\ref{noise_setup}(c) should be interpreted using
Eq.~(\ref{lesovikexp}).

\section{Relationship between noise and trapping-center spectral functions}

In this section we show that the TC noise spectrum Eq.~(\ref{tcnoise})
can be expressed as an integral over all possible
quasiparticle-quasihole excitations in the TC plus superconductor problem.
In order to derive this result, we define
the Matsubara and real time correlation functions as
follows\cite{fetter71}
\begin{subequations}
\begin{eqnarray}
{\cal S}(\tau-\tau')&=&-\Tr{\left\{
\hat{\rho}_{G}\hat{T}_{\tau}\left[
\delta\hat{n}(\tau)\delta\hat{n}(\tau')\right]\right\}},\label{stau}\\
S^{(R)}(t-t')&=&-i\theta(t-t')\Tr{\left\{
\hat{\rho}_{G}\left[\delta\hat{n}(t),\delta\hat{n}(t')\right]
\right\}},\quad\phantom{.}\label{srtau}
\end{eqnarray}
\end{subequations}
where we used the notation $\delta \hat{n}\equiv \hat{n}-2\bar{n}$.
Here, we use the Matsubara representation of operators 
$\hat{n}(\tau)=\textrm{e}^{\tau{\cal
    H}/\hbar}\hat{n}(0)\textrm{e}^{-\tau{\cal H}/\hbar}$, that are obtained
from the Heisenberg representation by substituting $it\rightarrow
\tau$.  

Applying Wick's theorem to Eq.~(\ref{stau}) leads to 
\begin{equation}
{\cal S}(\tau)=\sum_{\sigma,\sigma'}\left[{\cal G}_{\sigma\sigma'}(\tau){\cal
    G}_{\sigma'\sigma}(-\tau) -{\cal
    F}^{\dag}_{\sigma'\sigma}(\tau){\cal F}_{\sigma'\sigma}(-\tau)\right],
\label{wickstau}
\end{equation}
where we have introduced the normal ${\cal G}$ and anomalous ${\cal F}$ 
TC Matsubara Green's functions,
\begin{subequations}
\begin{eqnarray}
{\cal G}_{\sigma\sigma'}(\tau)&=&-\Tr{\{\hat{\rho}_G 
\hat{T}_{\tau}[d_{\sigma}(\tau)d^{\dag}_{\sigma'}(0)]\}},\label{greenG}\\
{\cal F}_{\sigma\sigma'}(\tau)&=&-\Tr{\{\hat{\rho}_G 
\hat{T}_{\tau}[d_{\sigma}(\tau)d_{\sigma'}(0)]\}}.\label{greenF}
\end{eqnarray}
\end{subequations}

We now take the Fourier transform of Eq.~(\ref{wickstau}),
$\tilde{S}(i\omega_n)=\int_{0}^{\beta\hbar}d\tau
\textrm{e}^{i\omega_{n}\tau}{\cal S}(\tau)$, 
and 
insert the Lehmann representation for the TC Green's functions,
\begin{subequations}
\begin{eqnarray}
{\cal
  G}_{\sigma\sigma'}(i\omega_n)
&=&
\hbar\int_{-\infty}^{\infty}d\omega'
\frac{{\cal A}_{\sigma\sigma'}(\omega')}
{i\omega_n-\omega'},\label{gleh}\\
{\cal F}_{\sigma\sigma'}(i\omega_n)
&=&
\hbar\int_{-\infty}^{\infty}d\omega'
\frac{{\cal B}_{\sigma\sigma'}(\omega')}{i\omega_n-\omega'}.\label{fleh}
\end{eqnarray}
\end{subequations}
The TC spectral functions ${\cal A}_{\sigma\sigma'}(\omega)$ and
${\cal B}_{\sigma\sigma'}(\omega)$ play a fundamental role in our
theory. For a BCS model such as Eq.~(\ref{htot}), we have
${\cal A}_{\uparrow\uparrow}={\cal A}_{\downarrow\downarrow}\equiv
{\cal A}$ with
${\cal A}$ real, and
${\cal A}_{\uparrow\downarrow}={\cal A}_{\downarrow\uparrow}=0$. Also, the
spectral function related to Gorkov's ${\cal F}$ function is non-zero
only for
${\cal B}_{\uparrow\downarrow}={\cal B}_{\downarrow\uparrow}\equiv
{\cal B}$, with ${\cal B}$ real. 
After inserting these Lehmann representations
into Eq.~(\ref{wickstau}), the result is readily evaluated using the
residue theorem, and taking advantage of the fact that
$\tilde{S}(i\omega_n)$ is non-zero only at even (Bose) Matsubara
frequencies [$\omega_{n}=n\pi/(\hbar\beta)$ with $n$ even]. 
Finally, analytic continuation ($i\omega_n\rightarrow \omega+i\eta$)
allows us to extract the TC noise spectrum from the imaginary part of
$\tilde{S}^{(R)}(\omega)$. 
This leads to a convenient expression for
the TC noise spectrum,
\begin{eqnarray}
\tilde{S}_{n}(\omega)&=&\hbar
\sum_{\sigma\sigma'}\int_{-\infty}^{\infty}d\epsilon'
\left[{\cal A}_{\sigma\sigma'}(\epsilon'){\cal A}_{\sigma\sigma'}
(\epsilon'-\omega)\right.\nonumber\\
&&\left.-{\cal B}^{*}_{\sigma'\sigma}(\epsilon')
{\cal B}_{\sigma'\sigma}(\epsilon'-\omega)\right]
[1-f(\epsilon')]f(\epsilon'-\omega).\quad\phantom{.}
\label{snsf}
\end{eqnarray}
Here the Fermi functions are given by 
\begin{equation}
f(\epsilon)=\frac{1}{\textrm{e}^{\beta\epsilon}+1}.
\end{equation}
The expression Eq.~(\ref{snsf}) is an exact result. Its derivation
relied on the use of Wick's theorem, which is valid only for a
quadratic Hamiltonian Eq.~(\ref{htot}) ($U=0$).\cite{fetter71}
It expresses the fact that the TC noise spectrum is
the sum of all quasiparticle-quasihole excitations involving the
dressed TC plus superconductor at thermal equilibrium.
Eq.~(\ref{snsf}) is the generalization of an equation derived by us
previously, using a canonical transformation in the TC plus normal
metal Fermi sea problem (See Eq.~(8) in Ref.~[\onlinecite{desousa05}]).

\begin{figure}
\includegraphics[width=3.in]{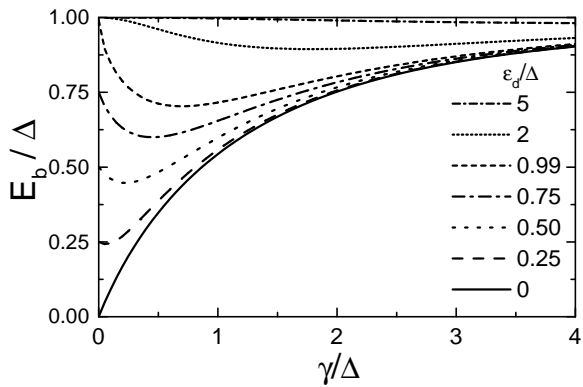}
\caption{Particle-like Andreev bound state as a function of 
  hybridization $\gamma$ for different trap energies $\epsilon_d$, all
  in units of the superconducting gap $\Delta$.  The hole-like Andreev
  bound state has the same energy but opposite sign.}
\label{andreev_levels}
\end{figure}

It is instructive to derive a sum rule for the noise spectrum starting
from Eq.~(\ref{snsf}). First, note that the TC occupation number and
TC pairing correlator are related to the spectral
functions in the following way,
\begin{subequations}
\begin{eqnarray}
\bar{n}_{\sigma}&=&\langle d^{\dag}_{\sigma}d_{\sigma}\rangle=
\int d\epsilon' {\cal A}(\epsilon')f(\epsilon'),\label{nsig}\\
F_{d}&=&-\langle d_{\uparrow}d_{\downarrow}\rangle=
\int d\epsilon' {\cal B}(\epsilon')f(\epsilon').\label{deltad}
\end{eqnarray}
\end{subequations}
Eq.~(\ref{deltad}) shows that the spectral function
${\cal B}(\epsilon')$ describes the extent to which TC pairing
is induced through its hybridization with the Fermi gas, i.e.,
${\cal B}(\epsilon')$ can be interpreted as a single-state proximity effect.
Integrating Eq.~(\ref{snsf}) over all frequencies and using
Eqs.~(\ref{nsig}),~(\ref{deltad}) we obtain the following sum rule,
\begin{equation}
\langle(\delta \hat{n})^2\rangle =
\int_{-\infty}^{\infty}\tilde{S}_{n}(\omega)d\omega
=\sum_{\sigma}\left[\bar{n}_{\sigma}\left(1-\bar{n}_{\sigma}\right)+F^{2}_{d}\right].
\label{sumrule}
\end{equation}
As a cross check, Eq.~(\ref{sumrule}) can be derived directly without
using Eq.~(\ref{snsf}) by simple applications of Wick's theorem and
use of the Fermi anticommutation relation, such that
$\hat{n}^{2}_{\sigma}=\hat{n}_{\sigma}$.  Interestingly, the sum rule
[Eq.~(\ref{sumrule})] shows that the onset of superconductivity tends
to \emph{increase} the amount of noise produced by a TC.

\section{Trapping-center spectral densities and Andreev bound states}

The TC spectral densities are known exactly for the case of zero
on-site Coulomb repulsion.\cite{yoshioka00,bauer07,hecht08} These can
be written as
\begin{subequations}
\begin{eqnarray}
{\cal A}&=&
  A\theta(|\epsilon|-\Delta)
  +\left[a_{+}\delta(\epsilon-E_b)
+a_{-}\delta(\epsilon+E_b)\right],\quad\phantom{.}\label{aform}
\\{\cal
  B}&=&B\theta(|\epsilon|
-\Delta) +b_{+}\left[\delta(\epsilon-E_b)
-\delta(\epsilon+E_b)\right],\label{bform}
\end{eqnarray}
\end{subequations}
where $\theta$ is the step function. Each spectral density is composed
of a continuous 
above-gap component which is non-zero only at
energies outside the superconducting gap ($|\epsilon|>\Delta$). For
energies within the gap, there are two sharp Andreev bound states,
with positive particle-like energy $E_b$ and negative
hole-like energy $-E_b$. These bound states are
reminiscent of the TC localized level $\epsilon_d$, whose energy is
renormalized to $\pm E_b$ due to hybridization with
Cooper pairs. 

In order to express these functions analytically, we define the trap 
hybridization parameter $\gamma$ as
\begin{equation}
\gamma=\pi g_0\langle V^{2}_{k}\rangle_{\epsilon_k=\epsilon_F},
\label{gamma}
\end{equation}
where $V_k$ is averaged over $\epsilon_k=\epsilon_F$ and $g_0$ is the
energy density at the Fermi level.
\begin{widetext}
The above-gap contributions are given by 
\begin{subequations}
\begin{eqnarray}
A(\epsilon)&=&\frac{\gamma|\epsilon|\sqrt{\epsilon^{2}-\Delta^2}
\left[\left(\epsilon+\epsilon_d\right)^{2}+\gamma^{2}\right]}
{\pi\left(\epsilon^{2}-\Delta^{2}\right)
\left[\left(\epsilon^{2}+\epsilon_{d}^{2}+\gamma^{2}\right)^{2}
-\left(2\epsilon\epsilon_{d}\right)^{2}
\right]+\left(2\epsilon\Delta\gamma\right)^{2}},\\
B(\epsilon)&=&\frac{-\Sgn{(\epsilon)}\gamma\Delta
\sqrt{\epsilon^{2}-\Delta^2}
\left[\epsilon^{2}-\epsilon_{d}^{2}-\gamma^{2}\right]}
{\pi\left(\epsilon^{2}-\Delta^{2}\right)
\left[\left(\epsilon^{2}+\epsilon_{d}^{2}+\gamma^{2}\right)^{2}
-\left(2\epsilon\epsilon_{d}\right)^{2}
\right]+\left(2\epsilon\Delta\gamma\right)^{2}},
\end{eqnarray}
\end{subequations}
\end{widetext}
where $\Sgn{(\epsilon)}$ denotes the sign of $\epsilon$.  We remark
that these are finite temperature spectral densities; the temperature
does not appear explicitly because the Matsubara Green's functions for
$U=0$ depend on temperature only through the Matsubara
frequencies.\cite{yoshioka00} The Andreev bound state energy is given
by the single pair of real roots $\pm E_b$ of
\begin{equation}
E^{2}\left(1+\frac{2\gamma}{\sqrt{\Delta^2-E^2}}\right) 
-\epsilon_{d}^{2}-\gamma^2=0,
\end{equation}
with the amplitudes $a_{\pm}$ and $b_{+}$ of
Eqs.~(\ref{aform})~and~(\ref{bform}) given by
\begin{subequations}
\begin{eqnarray}
a_{\pm}&=&\frac{\left(\Delta^2-E_b^{2}\right)\left[
\left(\epsilon_{d}\pm E_b\right)^{2}+\gamma^{2}
\right]}{2\left[\left(2\Delta^{2}-E_b^{2}\right)
\left(\epsilon_{d}^{2}+\gamma^{2}\right)-E_{b}^{4}\right]},\\
b_{+}&=&\frac{-\gamma\Delta E_b
\sqrt{\Delta^2-E_b^{2}}}
{\left[\left(2\Delta^{2}-E_{b}^{2}\right)
\left(\epsilon_{d}^{2}+\gamma^{2}\right)-E_{b}^{4}\right]}.
\end{eqnarray}
\end{subequations}
Note that $a_{+}\neq a_{-}$ in the asymmetric case $\epsilon_{d}\neq
0$, but $b_{+}=b_{-}$ always.  A useful relation is that
$b_+=-\sqrt{a_+a_-}$.  In Fig.~\ref{andreev_levels} we plot the
Andreev levels $E_b$ as a function of TC hybridization for different
TC energies $\epsilon_d$.

\begin{figure}
  \includegraphics[width=3.in]{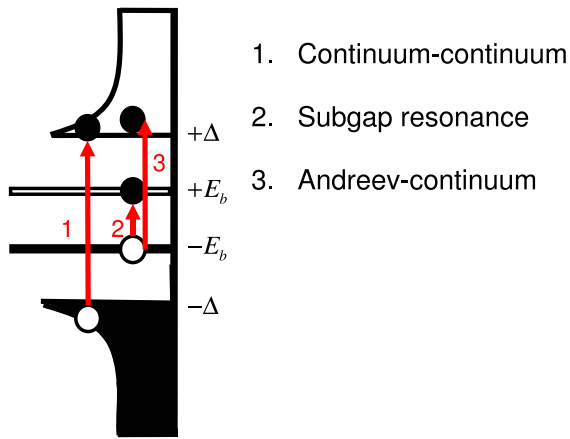}
  \caption{(Color online). Depiction of the energy density as a
    function of energy for the TC plus superconductor model, and the
    most important quasiparticle-quasihole excitations (denoted by
    arrows) determining the TC noise spectrum.\label{band_diagram}}
\end{figure}

It is useful to establish a connection to the case of a point contact
between superconductors.\cite{goffman00} In this case the transmission
of electrons across the point contact is dominated by the presence of
two Andreev bound states at equal and opposite energies with respect
to the Fermi level.  For zero phase difference these Andreev levels
are located close to $\pm \Delta$.  In our case, the trapping-center
is equivalent to a point contact provided $|\epsilon_{d}|\gg \Delta$;
Looking at Fig.~\ref{andreev_levels} we see that $E_b$
is indeed slightly below $\Delta$ in this limit. 

\section{Explicit results for the noise spectrum}

We now show explicit results for the noise spectrum of a single TC
hybridized with a superconductor. The analytic expressions for the
spectral functions are inserted into Eq.~(\ref{snsf}), where in the absence of a magnetic field
$\sum_{\sigma,\sigma'}= 2$ and ${\cal
  A}_{\sigma-\sigma}={\cal B}_{\sigma\sigma}=0$.
For $\omega\geq 0$ the noise is given by
\begin{widetext}
\begin{subequations}
\begin{eqnarray}
\tilde{S}_{n}(\omega)&=& 2\hbar
\left\{
\left[\int_{-\infty}^{-\Delta}+
\int_{\Delta+\omega}^{\infty} +
\theta(\omega-2\Delta)\int_{\Delta}^{-\Delta+\omega}\!\!\!\!\!d\epsilon
\right]
\left[A(\epsilon)A(\epsilon-\omega)
-B(\epsilon)B(\epsilon-\omega)\right]
\left[1-f(\epsilon)\right]f(\epsilon-\omega)\right. \label{contcont}\\
&&+2 a_{+}a_{-}
[1-f(E_b)]f(-E_b)\delta(\omega-2E_b)
\label{subgapres}\\
&&\!\!\!\!\!\!\!\!\!\!\!\!\!\!\!\!\!\!\!\!\left.+\sum_{\xi=+,-}
\theta(\omega-\xi E_b-\Delta) \left[a_{\xi} A
  (\xi E_b-\omega)+a_{-\xi}A(-\xi E_b+\omega) -2 b_{+} B
  (\xi E_b-\omega)\right] [1-f(\xi E_b)]f(\xi E_b-\omega)\right\}.\label{andcont}
\end{eqnarray}
\end{subequations}
\end{widetext}
The $\omega<0$ expression can be obtained from detailed balance
$\tilde{S}(-\omega) =\textrm{e}^{-\frac{\hbar\omega}{k_{B}
    T}}\tilde{S}(\omega)$.  

The positive frequency spectrum is interpreted as the sum over all
possible quasiparticle-quasihole pairs created when the TC plus Fermi
sea absorbs a photon with energy $\hbar\omega$ emitted by the noise
detector.  Fig.~\ref{band_diagram} illustrates the energy excitations
associated to TC noise.  The first contribution is a
continuum-continuum transition Eq.~(\ref{contcont}) where the hole
(particle) is in the continuum below (above) the superconducting gap.
This gives a smooth contribution to the noise spectrum when
$\omega>2\Delta$. The second line Eq.~(\ref{subgapres}) is the
\emph{subgap resonance}. The resonance occurs when a hole is created
at Andreev level $-E_b$ and a particle is excited at level $+E_b$.
This contribution is a sharp transition between Andreev levels: The
noise is a delta function peaked at $\omega=2E_b$.  The third line
Eq.~(\ref{andcont}) refers to transitions involving one of the Andreev
levels and the continuum.  This gives smooth contributions for
$\omega> \Delta\pm E_b$.

\begin{figure}
  \includegraphics[width=3.in]{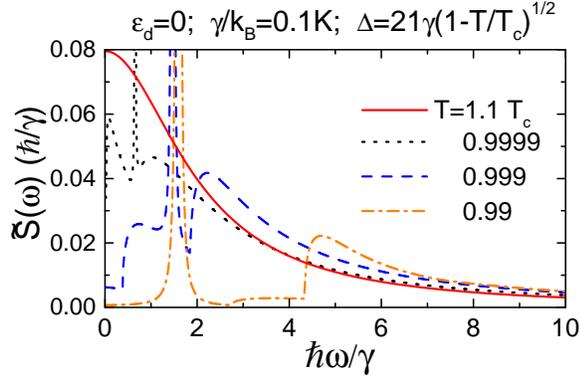}
  \caption{(Color online). Trapping-center noise spectrum near the
    superconducting transition temperature $T_c$, for $\epsilon_d=0$.
    For $T>T_c$ the noise has the Lorentzian form characteristic of
    random telegraph noise \cite{desousa05}. As $T$ is lowered below
    $T_c$ a gap opens in the noise spectrum, and a sharp subgap
    resonance appears as a transition between two Andreev bound states
    (for convenience, we represent the subgap resonance with a
    phenomenological linewidth equal to $0.01\gamma$). This shows that
    TC noise is dramatically affected by
    superconductivity.\label{noise_transition}}
\end{figure}

Fig.~\ref{noise_transition} shows the noise spectrum for temperatures
above and below the critical temperature for transition into the
superconducting state. We assume $\epsilon_d=0$, with the
superconducting energy gap dependent on temperature according to
$\Delta = 1.76 k_B T_c \sqrt{1-T/T_c}$ for $T\leq T_c$, and $\Delta=0$
for $T>T_c$.\cite{tinkham04} We assumed $k_B T_c/\gamma = 11.96$,
consistent with the value of $T_c=1.196$~K for aluminum with a trap
hybridization parameter $\gamma/k_B=0.1$~K.  For $T>T_c$ the noise is
a Lorentzian with linewidth $2\gamma/\hbar$, consistent with the high
temperature limit for random telegraph noise discussed in
Ref.~[\onlinecite{desousa05}] (Note that $k_BT_c\gg \gamma$ in
Fig.~\ref{noise_transition}). As the temperature is lowered below
$T_c$ a sharp resonance appears at energy equal to two 
$E_b$.To display the subgap resonance in the figure we
represented the delta function as a Lorentzian with linewidth equal to
$0.01\gamma$. 

For $T\ll T_c$, $k_BT\ll E_b$, and $\epsilon_d\lesssim
\Delta$ the noise is
well approximated by
\begin{eqnarray} 
  \tilde{S}_n(\omega)&\approx& 2\hbar
  \left\{2a_{+}a_{-}
    \delta(\omega-2E_b)+\theta(\omega-E_b-\Delta)\right.\nonumber\\
  &&\times\left[a_{+} A(E_b-\omega) +a_{-}A(-E_b +\omega)\right.\nonumber\\
&&\left.\left.-2b_{+} B(E_b-\omega)\right]\right\}.
\label{lowTnoise}
\end{eqnarray}

\begin{figure}
  \includegraphics[width=3.in]{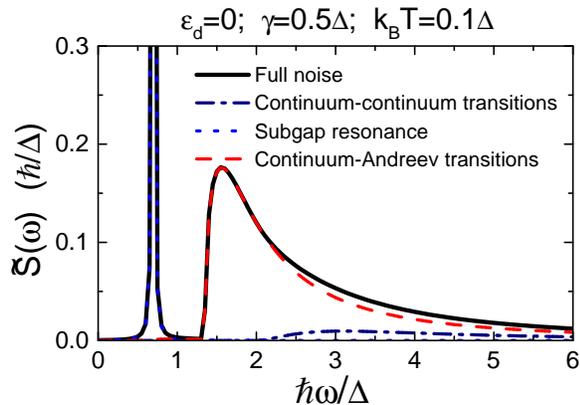}
  \caption{(Color online). Trapping-center noise for temperatures well below the
    superconducting transition. The parameters are $\gamma=0.5\Delta$,
    $\epsilon_d=0$, $k_B T =0.1\Delta$, leading to Andreev energy
    $E_b=0.35\Delta$.  About 60\% of the noise power is due to one
    sharp subgap resonance represented here by a Lorentzian with
    linewidth $0.001\Delta$. The remaining 40\% is dominated by
    processes involving the creation of a hole in the continuum and
    the excitation of an Andreev level at $+E_b$.  This occurs only
    for $\omega>\Delta+E_b$.}\label{noise_lowT}
\end{figure}

Fig.~\ref{noise_lowT} shows the low temperature noise spectrum ($T\ll
T_c$), with parameters normalized by the superconducting energy gap.
We also show the breakdown of 
the noise into its various contributions. For convenience, we
represented the subgap resonance as a Lorentzian with linewidth
$0.001\Delta$.  Our theory does not account for broadening mechanisms,
but we expect that disorder and other inhomogeneities will be a source
of broadening for Andreev bound states.  

For the parameters of Fig.~\ref{noise_lowT} the subgap
resonance accounts for $59$\% of the noise power.  The remainder is
due to Andreev-continuum transitions ($33$\%), with
continuum-continuum transitions contributing only $8$\%.
Remarkably, the continuum-continuum contribution is quite small, in
spite of being responsible for {\em all} the Lorentzian noise at $T>T_c$
(in the normal state). 

\begin{figure}
  \includegraphics[width=3.in]{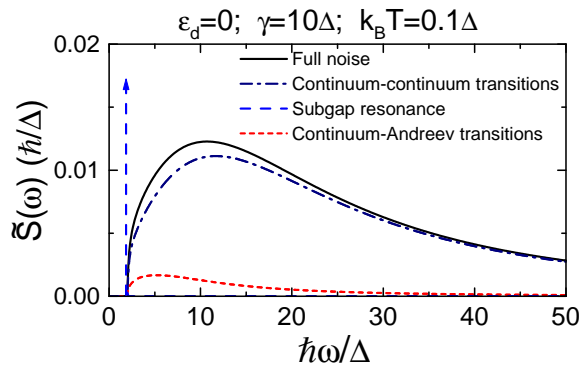}
  \caption{(Color online). TC noise spectrum in the superconducting regime, in a case
    where the Andreev bound states are very close to the gap edge
    ($E_b=0.981\Delta$). Here the noise is dominated by
    continuum-continuum contributions ($94$\% of the noise power),
    with the subgap resonance contributing only
    $0.3$\%.}\label{noisegamma10}
\end{figure}

Fig.~\ref{noisegamma10} shows the low temperature noise spectrum for a
case where the Andreev bound states are very close to the gap edge,
$E_b= 0.981\Delta$ (parameters $\epsilon_{d}=0$, $\gamma=10\Delta$,
and $k_BT=0.1 \Delta$). This case is quite different from
Fig.~\ref{noise_lowT}: The continuum-continuum contribution is now
$94$\% of the noise power, with Andreev-continuum contributions
$5.7$\%, and subgap resonance contributing only $0.3$\%.

\begin{figure}
  \includegraphics[width=3.in]{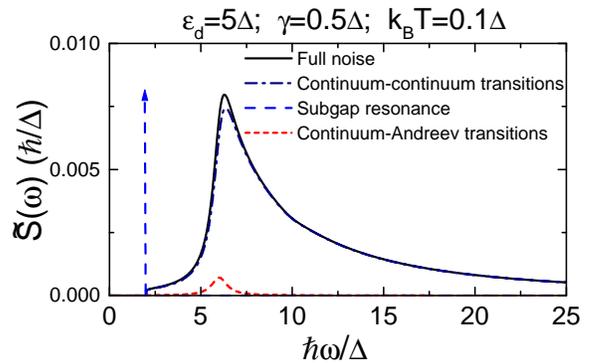}
  \caption{(Color online). TC noise spectrum for the asymmetric case
    $\epsilon_d=5\Delta$, with Andreev levels at $\pm E_b=\pm
    0.999\Delta$. The noise spectra has a smooth peak close to
    $\hbar\omega=6\Delta$. This occurs because the spectral function
    peaks at $\epsilon_d$. Similar to Fig.~\ref{noisegamma10}, the
    noise is dominated by continuum-continuum contributions ($98$\%)
    with the subgap resonance contributing less than
    $0.1$\%.}\label{noisegamma0p5}
\end{figure}

Fig.~\ref{noisegamma0p5} depicts the noise in the asymmetric regime
($\epsilon_d\neq 0$), with $\epsilon_{d}=5\Delta$, $\gamma=0.5\Delta$,
and $k_BT=0.1\Delta$ (the Andreev bound states are at
$E_b=0.999\Delta$).  Here the continuum-continuum contribution
accounts for $98$\% of the noise, with Andreev-continuum transitions
contributing $\approx 2$\% and subgap resonance contributing less than
$0.1$\%.  Interestingly, the noise has a broad peak at $\hbar\omega=6\Delta$, that
occurs because the spectral functions have a smooth peak at
$\epsilon_d=5\Delta$. 

Figs.~\ref{noise_lowT}-\ref{noisegamma0p5} show that the noise
  changes its character {\em completely} due to the opening of a gap and the
  formation of Andreev bound states in a superconductor.

\section{Andreev states as junction resonators \label{sectioneffh}}

We now relate our theory to the experimental observation of ``spurious
two-level systems'' (microresonators) in phase-based\cite{simmonds04}
and flux-based\cite{plourde05} superconducting qubits.

The model Hamiltonian for the interaction of a qubit with a TC plus
Fermi sea is simplified by projecting onto the Hilbert space of
Andreev bound states. This is achieved by expressing the TC operators as
\begin{subequations}
\begin{eqnarray}
  d_{\uparrow}&=&u\left(\frac{\alpha^{\dag}_{-}+\alpha_{+}}{\sqrt{2}}\right)
  +v\left(\frac{\alpha_{-}-\alpha_{+}^{\dag}}{\sqrt{2}}\right)+d_{\uparrow,{\rm cont}},\quad\phantom{.}\label{dup}\\
  d_{\downarrow}^{\dag}&=&-v\left(\frac{\alpha^{\dag}_{-}+\alpha_{+}}{\sqrt{2}}\right)
  +u\left(\frac{\alpha_{-}-\alpha_{+}^{\dag}}{\sqrt{2}}\right)+d_{\downarrow,{\rm cont}},\qquad\phantom{.}\label{ddown}
\end{eqnarray}
\end{subequations}
where $\alpha^{\dag}_{\pm}$ is a creation operator for an Andreev
level with energy $\pm E_b$, and the $d_{\sigma,{\rm cont}}$ denote the additional
operators acting on the continuum. The canonical transformation defined by
Eqs.~(\ref{dup})~and~(\ref{ddown}) diagonalizes our TC model when
$u=\sqrt{a_+}$ and $v=\sqrt{a_-}$. This can be verified by calculating
the Green's function using the canonical transformation and comparing
to Eqs.~(\ref{aform})~and~(\ref{bform}). Substituting
Eqs.~(\ref{dup})~and~(\ref{ddown}) into Eq.~(\ref{qubitHtrans}) we get an effective Hamiltonian for the qubit
interacting with a pair of Andreev bound states,
\begin{eqnarray}
{\cal H}_{\rm{Q-A}}&=&\frac{1}{2}\hbar\Omega_0 \hat{\sigma}_z 
+E_b \alpha_{+}^{\dag}\alpha_{+}-E_b\alpha_{-}^{\dag}\alpha_{-}\nonumber\\
&&+ \left(\lambda_z \hat{\sigma}_z +\lambda_x \hat{\sigma}_x\right) \left[
2\sqrt{a_+a_-} \left(\alpha_{+}^{\dag}\alpha_{-}+\alpha_{-}^{\dag}\alpha_{+}\right)\right.\nonumber\\
&&\left.+(a_{+} - a_{-})\left(\alpha_{+}^{\dag}\alpha_+ -\alpha_{-}^{\dag}\alpha_{-}\right)
\right].
\label{effh}
\end{eqnarray}
Here $\lambda_z=\hbar(\delta I_c)\eta_z$ and $\lambda_x=\hbar(\delta
I_c)\eta_x$ are characteristic coupling energies between the qubit and
the Andreev levels. For phase qubits these should be a fraction of the
change in Josephson energy $\hbar(\delta I_c)/(2e)$.  Recall from
section~\ref{section_qubit} that the $\eta$'s depend on qubit
design, while $(\delta I_c)$ is the characteristic shift in
critical current due to a TC. A similar expression will hold for other
kinds of qubits, for example, in a Cooper-pair box $\lambda_i \sim
p_{\rm{Q}}p_{\rm{TC}}/R^{3}$ is the electrostatic energy due to the
interaction of the Qubit's electric dipole moment $p_{Q}$ and the TC
(dipole moment $p_{\rm{TC}}$ due to the image charge produced at the
reservoir).\cite{desousa05}

The qubit-Andreev interaction is weighted by additional factors
accounting for the branching of the impurity spectral weight into
different channels --- not all of the impurity's noise goes into the
Andreev channel. The first interaction,
$2\sqrt{a_+a_-}(\alpha_{+}^{\dag}\alpha_{-}+\alpha_{-}^{\dag}\alpha_{+})(\lambda_z
\hat{\sigma}_z+\lambda_x \hat{\sigma}_x)$ produces admixture between
qubit and Andreev levels, and leads to important anticrossings in
qubit spectrometry. The second interaction,
$(a_+-a_-)(\alpha_{+}^{\dag}\alpha_+
-\alpha_{-}^{\dag}\alpha_{-})(\lambda_z \hat{\sigma}_z+\lambda_x
\hat{\sigma}_x)$ only exists in the asymmetric case ($\epsilon_d\neq
0$). It enables the design of quantum gates through electrical
manipulation of Andreev states.

\begin{figure}
\includegraphics[width=3.in]{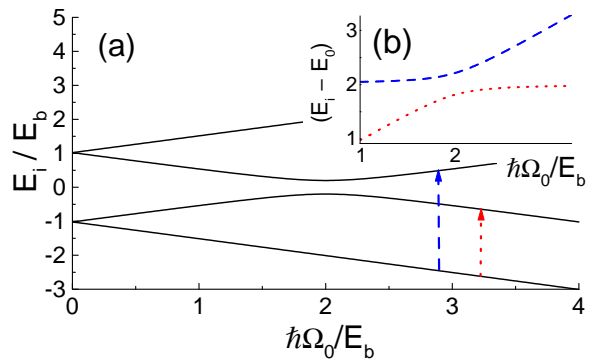}
\caption{(Color online). (a) Energy level structure of the effective Hamiltonian 
  for a superconducting qubit interacting with a pair of Andreev bound
  states (one hole-like at energy $-E_b$ and one
  particle-like at $+E_b$). The Andreev-qubit coupling is
  assumed $\lambda=0.2E_b$.  Anticrossing behavior occurs
  when the qubit energy splitting matches the subgap resonance,
  $\hbar\omega=2E_b$. (b) Energy differences between the
  ground state and the first and second excited states. The result is
  remarkably similar to spectroscopy measurements on
  phase\cite{simmonds04} and flux qubits.\cite{plourde05}}\label{two_levels}
\end{figure}

The Hamiltonian Eq.~(\ref{effh}) describes a four-level system,
where the qubit energy levels are hybridized with the pair of Andreev
states; it serves as a starting point to study non-equilibrium effects
for a qubit coupled to Andreev excitations.  Fig.~\ref{two_levels}(a)
shows the energy levels $E_i$ obtained after diagonalizing
Eq.~(\ref{effh}) for $\lambda_x=0.2E_b$ and $\lambda_z=0$. Note the
level anticrossing when $\hbar\Omega_0=2E_b$.  Fig.~\ref{two_levels}(b)
shows the two lowest energy transitions measured by qubit
spectroscopy, $E_1-E_0$ and $E_2-E_0$. We remark the similarity of our
Fig.~\ref{two_levels}(b) to the experimental data in Fig.~2(a) of
Ref.~[\onlinecite{simmonds04}]. For these frequencies the qubit is
highly mixed with the Andreev excitation.

Therefore each pair of Andreev levels acts as a microresonator, with
frequency in the range $2E_b \in (0,2\Delta)$.  The anticrossing behavior
occurs only when the qubit is in resonance with a transition between
Andreev levels, i.e., 
when the qubit frequency coincides with a subgap
resonance in the noise spectrum.

\section{Discussion and conclusion\label{discussionandconclusions}} 

In summary, we calculated the noise spectrum due to individual
trapping-centers (TCs) hybridized with a superconducting lead.  We
showed that the opening of a gap and the formation of Andreev bound
states change the character of the noise completely.  At $T<T_c$, the
noise is substantially different from the usual Lorentzian spectra
assumed in simple models.

In many cases the noise is dominated by a subgap resonance related to
transitions between Andreev bound states at energies $\pm E_b$
reminiscent of the localized TC states.  At $T\ll T_c$, the subgap
resonance may account for over half of the noise power (See
Fig.~\ref{noise_lowT}).  The remaining noise power occurs only at
$\hbar\omega>\Delta +E_b$, giving a smooth gapped spectrum related to
the excitation of an Andreev level into the continuum.

We assumed a TC model with zero on-site Coulomb repulsion.  As a
result, the noise can be expressed exactly as an integral over TC
spectral densities, which are known analytically. 
This constitutes a limiting case which provides a fully characterized reference point. We now discuss the
expected role of TC Coulomb repulsion. Spectral densities for $U>0$
were calculated using the numerical renormalization group method in
Refs.~[\onlinecite{bauer07,hecht08}]. For $U>0$, the energy $E_b$ of
each Andreev level is shifted, but the number of Andreev levels
remains the same (one hole-like and one particle-like per
TC).\cite{bauer07,hecht08}

At $U=0$, the total noise power [Eq.~(\ref{sumrule})] is appreciable
only if $\epsilon_d$ lies within the interval $[-\Delta,\Delta]$ (or
within $\Max \{ k_BT,\gamma \}$ of this interval. For $T>T_c$ this
result is equivalent to the one found in Ref.~[\onlinecite{desousa05}]).
An interesting open question is whether this result will change for
$U>0$.

We derived an effective Hamiltonian for a superconducting qubit
interacting with a TC, showing that the qubit sees the TC as two
Andreev levels. Anticrossing occurs when the qubit frequency is in
resonance with the energy separation of the two Andreev levels.  This
gives a microscopic explanation for the experimental observation of
microresonators coupled to Josephson-junction devices. Simmonds and
collaborators observed anticrossing behavior at a number of
frequencies in the spectroscopy of Josephson-junction phase
qubits.\cite{simmonds04,cooper04} Plourde and collaborators observed a
similar effect in the spectroscopy of flux qubits.\cite{plourde05} Kim
{\it et al.} \cite{kim08} observed avoided level crossings in the
spectra of a Cooper-pair box.

Our work
establishes a direct connection between TCs in the Josephson-junction
insulator and the presence of these anticrossings.

Another interesting implication of our model is that each TC will
become a sharp dielectric resonance \emph{only} when the lead becomes
a superconductor. TCs are charged defects, possessing an electric
dipole moment due to their image charge in the superconducting lead.
The fluctuation-dissipation theorem implies that the power absorbed by
a TC irradiated by an AC electric field at frequency $\omega$ is given
by $P_{\omega}\propto\omega \tilde{S}_{n}(\omega)$. Therefore, the
subgap resonance in $\tilde{S}_{n}(\omega)$ can be detected as a sharp
resonance in dielectric absorption $P_\omega$ (in the normal state,
$P_\omega$ will be a broad resonance, see
Fig.~\ref{noise_transition}).

This effect provides a powerful method to validate our theory
experimentally. There are two other microscopic models for the
microresonator: Macroscopic resonant tunneling\cite{johnson05} results
in no dielectric resonance; structural two-level
system\cite{martinis05,constantin07} gives rise to the \emph{same}
dielectric resonance above and below $T_c$. Hence microwave absorption
experiments above and below $T_c$ will clearly reveal whether the
microresonator is a pair of Andreev levels or not.

In conclusion, we 
have developed a microscopic theory for critical current
and charge noise in superconducting devices based on a charge
tunneling model with individual trapping-centers. We showed that the
superconducting gap and the formation of Andreev levels plays a
prominent role in determining the noise spectrum, providing a
microscopic explanation for the microresonators observed in
experiments. Our calculated noise spectrum is drastically different
from the usual phenomenological Lorentzian and $1/f$ noise 
spectra derived in previous work.

\acknowledgements

RdS and FKW acknowledge support from NSERC-Discovery. RdS also
acknowledges support from the University of Victoria Faculty of
Sciences. TH and JvD acknowledge support from the DFG through SFB631,
SFB-TR12, and the German Excellence Initiative via the ``Nanosystems
Initiative Munich (NIM)'', as well as partial support from DIP-H.2.1,
and from the NSF under grant No. NSF PHY05-51164.

\end{document}